\documentclass[a4paper, conference]{IEEEtran}
\IEEEoverridecommandlockouts
\usepackage{cite}
\usepackage{amsmath,amssymb,amsfonts}
\usepackage{algorithmic}
\usepackage{graphicx}
\usepackage{textcomp}
\usepackage{subcaption}
\usepackage{xcolor}
\usepackage{float}
\usepackage{tabularx} % Importa o pacote tabularx
\usepackage{lscape} % Pacote para orientação em paisagem
\usepackage[left=1.62cm,right=1.62cm,top=1.9cm]{geometry}

\begin{document}

\title{Performance Modeling and Evaluation of Hyperledger Fabric: An Analysis Based on Transaction Flow and Endorsement Policies}

\author{\IEEEauthorblockN{Carlos Melo, Glauber Gonçalves, Francisco A. Silva, and André Soares}
\IEEEauthorblockA{Universidade Federal do Piauí (UFPI), Picos - PI, Brazil\\
}
\IEEEauthorblockA{\{casm, ggoncalves, faps, andre.soares\}@ufpi.edu.br}
}

\maketitle

\begin{abstract}
Blockchain is a paradigm derived from distributed systems, protocols, and security concepts. 
However, can blockchain applications provide services in industrial environments, especially concerning performance issues?
In blockchains, long response times can impair both user and service experience, and intensive resource use may increase the costs of service provision.
The proposed paper tries to answer this question by evaluating the performance of one of the most popular permissioned blockchain platforms, the Hyperledger Fabric (HLF).
We provide a framework for performance evaluation based on modeling and experimentation.
The results indicate that block size and arrival rate can compromise throughput (by -70\%), latency (by +1,500\%), and environment utilization (by +28\%) and that multiple gateways can reduce latency (by -75\%), and throughput (by -60\%).
\end{abstract}


\section{Introduction}

Hyperledger Fabric (HLF) is an open-source permissioned blockchain platform that has been evaluated in various industry and corporate scenarios~\cite{androulaki2018hyperledger}.
Unlike public blockchains, such as Bitcoin and Ethereum, based on anonymous participation, HLF focuses on the network among entities that share common goals but may lack full trust in each other due to business competition. Thus, this blockchain platform requires the identification of every participant in the network organized into private communication channels.

HLF offers identifiable nodes, varying security levels for transaction execution with endorsement policies, and an architecture that executes-orders-validates transactions with modular ordering for both Crash Fault Tolerance (CFT) and Byzantine Fault Tolerance (BFT) models\cite{androulaki2018hyperledger}.
The HLF perform better than Bitcoin and Ethereum's order-execute-model while bolstering security \cite{thakkar2018performance}.

Typically, transactions carried out on permissioned networks are not linked to consumption of \textit{gas} or monetary cost.
There is no charge for a transaction to be executed, which facilitates its use in private environments\cite{antwi2021case}, requiring higher throughput and lower latency at the lowest possible financial cost.
However, quantifying the resources necessary for these applications' provision and full utilization has become an important task related to the performance evaluation process.

Performance in the context of blockchain is a key factor used to demonstrate the viability of adopting this technology.
It has become a relevant research topic in recent years, with studies focused on performance modeling and measurement~\cite{xu_ipm2021, thakkar2018performance, guggenberger2022depth, sukhwani2018performance, melo_computing2022, wu_acm_ease2022, jiang2020performance, ke_springer_cc2022, yuan2020performance}.
However, due to the volatility of this research field, most works need to consider the new versions and readjustments introduced in the context of these platforms and how these changes impact the service provisioning performance.

Also, some of these works, like \cite{silva2023avaliaccao}, do not provide a model validation, while others, such as \cite{thakkar2018performance}, do not provide generalizable performance results, attaining it to their experimental environment.
This paper uses measurement and modeling to compose a generalizable performance evaluation \textit{framework} within the scope of HLF 2.5 and beyond.

This work offers two main contributions: (i) a generalized performance model representing the HLF transaction flow and (ii) a performance evaluation through case studies to validate and demonstrate the feasibility of the proposed model.
The results allow HLF service providers to adjust parameters according to their infrastructure, focusing on \textit{throughput}, latency, and resource utilization.
Furthermore, we evaluate the impact of endorsement policy and distribute endorsement on these metrics, enabling organizations that do not share mutual trust to enjoy the benefits of blockchains.

The next sections are organized as follows:
Section \ref{sec:related} presents related works in modeling and performance evaluation of blockchains.
Section \ref{sec:proposed} introduces the base model and discusses its applications.
Section \ref{sec:architecture} provides an overview of the transaction flow within the Hyperledger Fabric platform.
Section \ref{sec:measurement} describes the experimental study and the validation process of the proposed model for the HLF platform.
Section \ref{sec:results} presents a set of case studies and derivations of the proposed model, as well as the main results obtained in each evaluation scenario.
Finally, Section \ref{sec:conclusion} summarizes our final considerations, limitations, and future works.

%---------------------------------------------------------------------------------

\section{Related Works}
\label{sec:related}

The Hyperledger Fabric platform has been extensively discussed in previous research, focusing on performance metrics.
This paper seeks to update the state of the art and fill gaps left by existing studies.

In \cite{melo_computing2022}, we introduced models to assess the use of computational resources in the context of HLF.
Using Continuous Time Markov Chains (CTMCs) and Stochastic Petri Nets (SPNs), we demonstrated the effectiveness of these formalisms in modeling and evaluating HLF-based applications, particularly in detecting infrastructure bottlenecks.
However, the model proposed in this paper extends the scope of the previous work, assessing metrics such as throughput and transaction latency not addressed by previously proposed models.

In other studies like \cite{wu_acm_ease2022, jiang2020performance, ke_springer_cc2022}, the authors modeled general system performance metrics from the perspective of HLF.
Jiang et al.\cite{jiang2020performance}, for example, used a hierarchical modeling approach for HLF 1.4, analyzing indicators such as throughput, latency, and system utilization.
Whereas Wu et al.\cite{wu_acm_ease2022} developed a queue theory-based model focused on the flow of a transaction within the scope of HLF 2.0.
Ke and Park~\cite{ke_springer_cc2022} proposed queue models considering different service rates.
% However, these studies did not include a formal bottleneck detection method, such as applying sensitivity analysis to quantify the impact caused by each platform component on the metric of interest.

Finally, in other works like \cite{sukhwani2018performance, yuan2020performance}, the authors evaluated the influence of arrival rates and block sizes on HLF throughput and latency. Still, as we will see in the present work, these are not the only factors significantly impacting these metrics.
Furthermore, in Yuan et al.\cite{yuan2020performance}, the authors modeled the platform using Generalized Stochastic Petri Nets (GSPN). In contrast, in Sukhwani et al.\cite{sukhwani2018performance}, the authors used Stochastic Reward Networks (SRN); both models are isomorphic to the SPNs presented in this work, that is, they have the same symbolic power; however, they do not contemplate the transaction flow for more recent versions of HLF.

This work extends those studies, especially directly extends our previous work \cite{silva2023avaliaccao}, by presenting an SPN model for the Hyperledger Fabric platform, focusing on the HLF 2.5+ version addressing the concept of the gateway and the possibility of it being a single point of failure and a pre-platform bottleneck.
% The methodology proposed in developing this work includes a formal validation and bottleneck identification in the proposed model, increasing the practical relevance of the findings through case studies.

%---------------------

\section{Architecture}
\label{sec:architecture}

\begin{figure}[htpb]
\centering
\includegraphics[width=.48\textwidth]{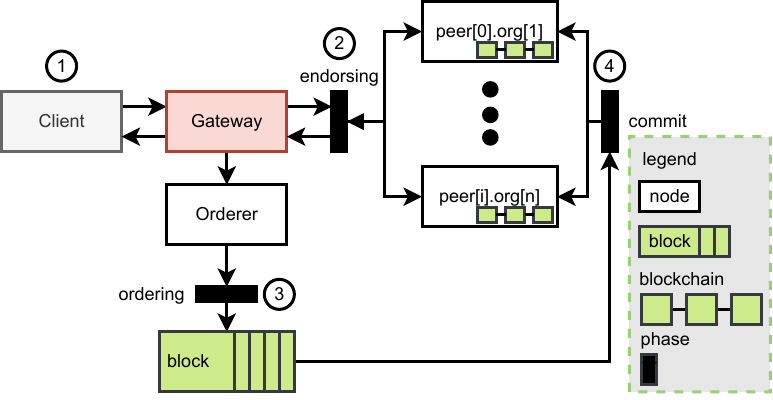}
\caption{Hyperledger Fabric - Transaction Flow}
\label{fig:overview}
\end{figure}

Figure \ref{fig:overview} demonstrates the transaction flow in the network through a process subdivided into four stages: \textit{proposal}, \textit{endorsement}, \textit{ordering}, and \textit{commit}. 
The figure also shows network components, such as the client and gateway, where transactions are submitted to the blockchain, and peers that must maintain consistent replicas of the blockchain.%%

1) From version 2.5 of HLF, the flow begins with the client application sending a transaction proposal to a \textit{gateway}, which acts as an intermediary between the client and the peers;
2) Endorsing peers will simulate the transactions and determine if they meet the pre-established requirements in the application's smart contract, which is a self-executing contract with business rules directly written into code;
3) The transaction is then forwarded to the ordering peers (\textit{orderer}), who will insert it into a block together with other transactions;
4) The generated block is subsequently persisted on the blockchain through the \textit{commit} process.

A transaction in HLF consists of read and write operations that change the values of data objects in the blockchain. Transactions can range from creating a new data object to transferring the account owner of objects.
In turn, the endorsing phase checks whether transaction operations lead to a consistent state of the blockchain by just simulating a given transaction.
For example, it is common for the simulation to verify whether the sender indeed owns the goods and that the recipient exists in a transaction transferring goods between accounts.
It is also common for endorsing peers to belong to different geographically distributed organizations (\textit{org}) that do not have mutual trust in each other.

The endorsed transaction returns to the gateway, and it will be forwarded to the orderers responsible for managing new blocks.
Blocks have a predefined size and a maximum wait time to be formed (\textit{batch timeout}).
Once transactions are ordered into a block, this block will be sent to the committers nodes, which will process and insert it into their replicates of the blockchain to ensure integrity, consistency, and immutability of data objects. 
Note that endorsing and commit phases can execute in the same network node, as shown in Figure~\ref{fig:overview}, or be separated into distinct nodes depending on the blockchain governance model.

%----------------------------------------

\section{Proposed Model}
\label{sec:proposed}

\begin{figure*}
    \centering
    \includegraphics[width=.73\textwidth]{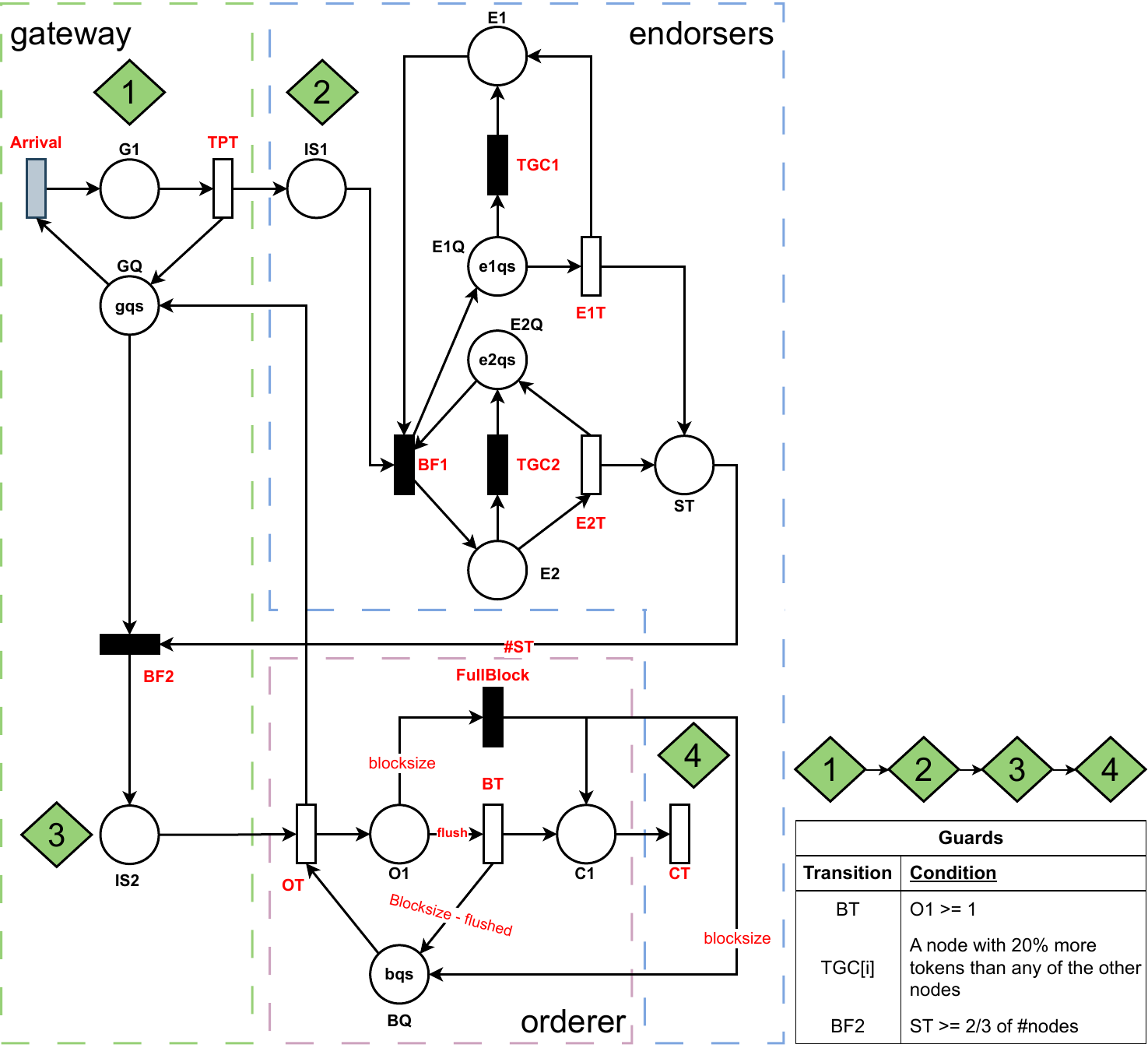}
    \caption{Proposed Model - Hyperledger Fabric}
    \label{fig:figura}
\end{figure*}

Figure \ref{fig:figura} presents the proposed model based on the high-level transaction flow of the HLF platform defined in Figure \ref{fig:overview}.
All transitions in the model follow single-server semantics and exponential distribution with distinct times.

Transaction flow in the model begins with triggering the \textbf{Arrival} transition, depicted in gray at the top left of the model.
The first highlighted area composed by a rectangle formed by dashed green lines emphasizes the activities occurring within the gateway.

Transactions arriving through Arrival are directed to the \textbf{G1} location, symbolizing the \textit{gateway} in the HLF environment, with G1 connected to the \textbf{GQ} (\textit{gateway queue}) location through the \textbf{TPT} (\textit{transaction posted}) transition.
Transactions exceeding the \textit{gateway} capacity are discarded.
Transactions entering the gateway proceed to the intermediate state \textbf{IS1} via the timed \textbf{TPT} transition and are directed towards the endorsing peers through the \textbf{BF1} transition.

The endorsing peers and their respective capacities are represented by the locations \textbf{E1}, \textbf{E2}, \textbf{E1Q}, and \textbf{E2Q}.
A garbage collector mechanism, \textbf{TGC}, is present in all endorsing peers, reducing the probability of transaction \textit{starvation} within the system and the waste of computational resources generated by software aging.

After endorsement, the transaction reaches the \textbf{ST} (\textit{Stamped Transaction}) state through the average endorsement time represented by the \textbf{E1T} and \textbf{E2T} transitions. A transaction in the modeled environment receives endorsement from two peers, meaning the \textbf{ST} location will have up to two markings representing a single transaction. The attested transaction waits for the gateway's availability to receive it (\textbf{BF2}) and then moves to the intermediate state \textbf{IS2} via the \textbf{BF2} branching transition, indicating that the transaction is between the gateway and the orderer.

The \textbf{OT} (\textit{ordering time}) transition represents the time to order transactions into a new block through the ordering peer (\textbf{O1}). The \textbf{BQ} (\textit{block queue}) location indicates the block size or the maximum number of transactions required to complete it \cite{silva2023avaliaccao}.

In HLF, a block can be partially or filled. Upon reaching the full capacity of a block, transactions in \textbf{OQ} move to the \textbf{C1} location through the \textbf{FullBlock} transition. The alternative flow used for a partially filled block goes through the \textbf{BT} (\textit{Batch Timeout}) location, which assists in preventing \textit{starvation} by establishing the maximum time a transaction will be in the ordering phase.

The \textbf{flush} variable on the arc connecting the \textbf{O1} location to the \textbf{BT} transition ensures the transfer of transactions to \textbf{BQ} without exceeding the block size (\textit{blocksize}). Blocks enter the persistence phase through the \textbf{C1} location via the \textbf{CT} transition. New blocks are \textit{committed} to the other peers. 
For better accuracy, guard expressions in \textbf{BF2}, \textbf{BT}, and \textbf{TGCs}, which are presented in Figure \ref{fig:figura}, are responsible for restricting the firing of these transitions, ensuring the model's efficiency and representativeness when compared to the real environment.

%--------------------------------------------
\section{Experimental Methodology and Validation}
\label{sec:measurement}

The experiments to validate the base model were conducted using the basic network of Hyperledger Fabric.
The basic network, maintained by the HLF developers, enables user creation, verification, and transfer of assets.
The focus of the experiments was exclusively on asset creation transactions due to their higher resource demand than querying operations and transferring goods between accounts.

The setup of the experimental environment included a full node with two endorsing peers and one orderer deployed in Docker containers.
The test machine had four physical cores, 8 GB of RAM, and 80 GB of storage.
The software consisted of Ubuntu 22.04, Hyperledger Fabric 2.5, and Docker 24. The orderer, also a Docker container, organized transactions into blocks and ensured their distribution and persistence in the network.
A command-line interface (CLI) in a separate container facilitated the interaction between the client application and the blockchain network through the container containing the gateway.

In the validation process of the proposed model, the client application, developed in TypeScript, sent an average of 10 transactions per second (tps) to the platform's gateway. The client sent the transactions to the system following an exponential distribution, the same adopted for the timed transitions of the proposed model.

The goal of the experiments was to validate the model, not to optimize resource usage through high demand that would result in an overload on the peers.
However, we will see in our case studies that many components were on the verge of resource exhaustion, even with a low arrival rate.
Thus, the transaction arrival rate was adjusted to ensure the execution of all phases in the model, including the times required for forming partial and complete blocks.

Measurements were carried out in the environment, and the obtained data were used to evaluate the proposed model.
% Figure \ref{fig:comparison} illustrates the obtained results, displaying latency from measurements and simulations.
% \begin{figure}[htpb]
% \centering
% \includegraphics[width=0.48\textwidth]{img/validation_area_plot.pdf}
% \caption{Comparison between Measured and Expected Latency}
% \label{fig:comparison}
% \end{figure}%
The experimental evaluation involved submitting 400 transactions to the real system. The average times obtained for each stage were then used to validate the proposed model through 400 simulations.
The Mercury tool \cite{silva2015mercury} was used for the simulations, which focused on latency as the key metric.
The measured results indicated an average latency of 1131 ms with a standard deviation of 408 ms.
An average latency of 1110 ms with a standard deviation of 18 ms for the simulated results.

A two-sample t-test was conducted, resulting in a p-value of 0.3.
With a significance level $\alpha$ of 0.05, the test indicated no statistically significant difference between the results obtained through model evaluation and experiments.
Therefore, with 95\% confidence, there is no evidence to reject the hypothesis that the proposed base model accurately represents the system.

%--------------------------------------------
\section{Case Studies}
\label{sec:results}

% After validating the proposed model, we focus on specific scenarios and establish a set of case studies to evaluate critical performance metrics such as latency, throughput, and utilization in blockchain environments.
% Moreover, we assess the impact of applying endorsement policies on the platform's overall performance and how geographically distributed endorsement impacts the relationship between different organizations.

The reference data for the model are the average times for each factor obtained experimentally through measurement. 
Table \ref{tab:variation} summarizes these values, which will serve as input parameters in the model.

\begin{table}[htpb]
\caption{Input Values for the Model and Their Variations}
\footnotesize
\centering
\begin{tabular}{lrr}
\hline
\multicolumn{1}{c}{\textbf{Factor}} & \multicolumn{1}{c}{\textbf{Baseline}} & \multicolumn{1}{l}{\textbf{Range \{min,max\}}} \\ \hline
Arrival Rate ($\lambda$)       & 10/s  & \{5, 18\} \\
Block Timeout (BT)       & 2000 ms & \{1000, 3000\} \\
Endorsement Time (E1T and E2T)     & 160 ms & \{80, 240\}  \\
Ordering Time (OT)        & 15ms   & \{7.5, 22.5\}  \\
Commit Time (CT)       & 1150 ms  & \{575, 1725\}    \\
Transaction Posting Time (TPT)          & 10 ms    & \{5, 15\}      \\
Capacity Sizes (E1Q, E2Q, GQ) & 10 & \{5, 15\}      \\
Block Size (BQ) & 10 & \{5, 50\}      \\ \hline
\end{tabular}
\label{tab:variation}
\end{table}

\subsection{Latency, Throughput, and Utilization}

The first analyzed metric is latency, derived from Little's Law \cite{jain1991}, which describes the relationship between the average number of pending transactions in a system and its respective arrival rate.
We calculate the latency for the proposed model according to Equation \eqref{eq:expected}, where \( P(m(\text{Place})=i) \) is the probability of \( i \) markings in the place.

\begin{equation}
\label{eq:expected}
E(\text{Place}) = \sum_{i=1}^{n} P( m(\text{Place})=i) \times i,\\
\end{equation}

\begin{figure*}[htbp]
  \centering
  \begin{subfigure}[b]{0.32\linewidth}
    \centering
    \includegraphics[width=\linewidth]{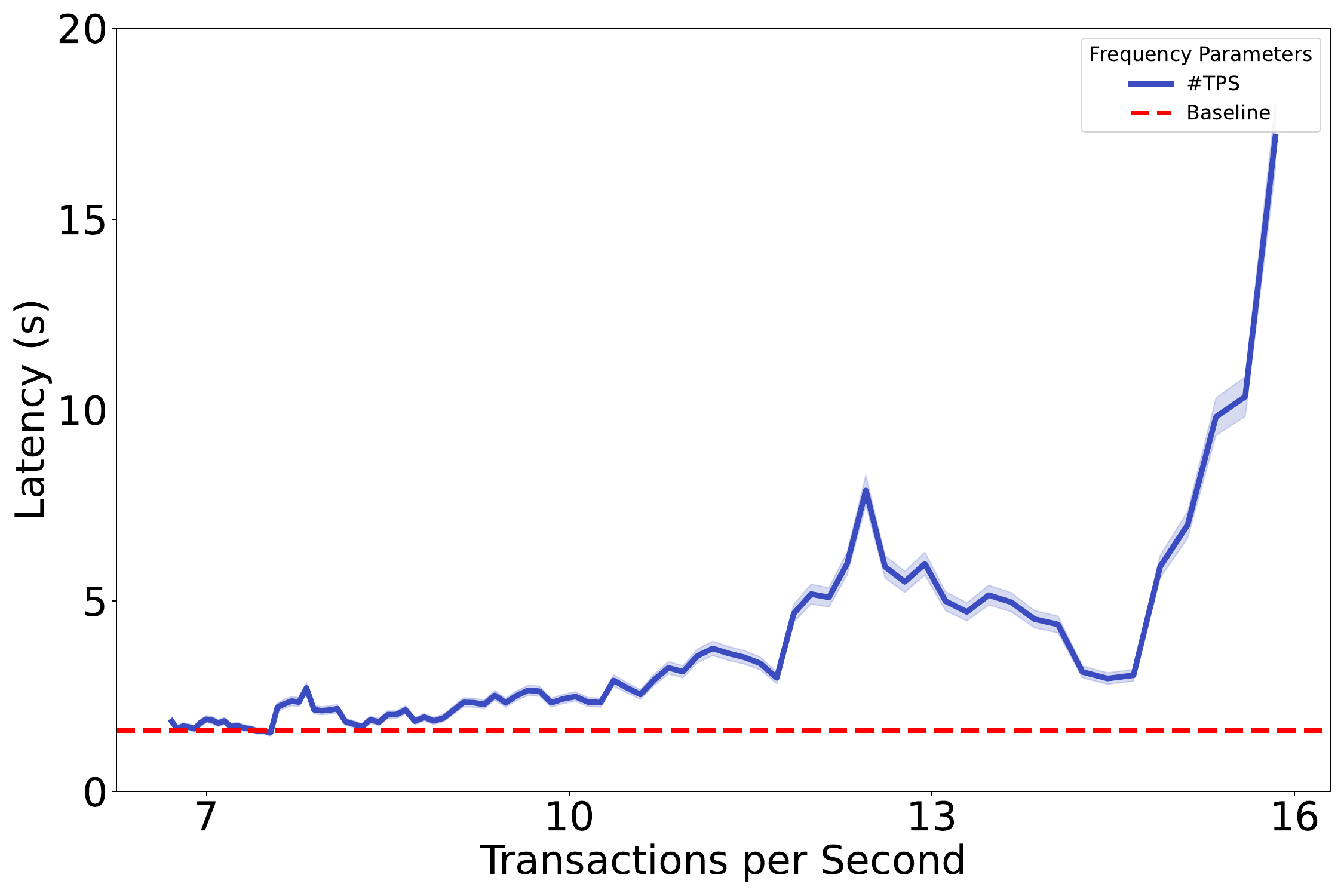}
    \caption{}
    \label{fig:arrival_mrt}
  \end{subfigure}
  \hfill
  \begin{subfigure}[b]{0.32\linewidth}
    \centering
    \includegraphics[width=\linewidth]{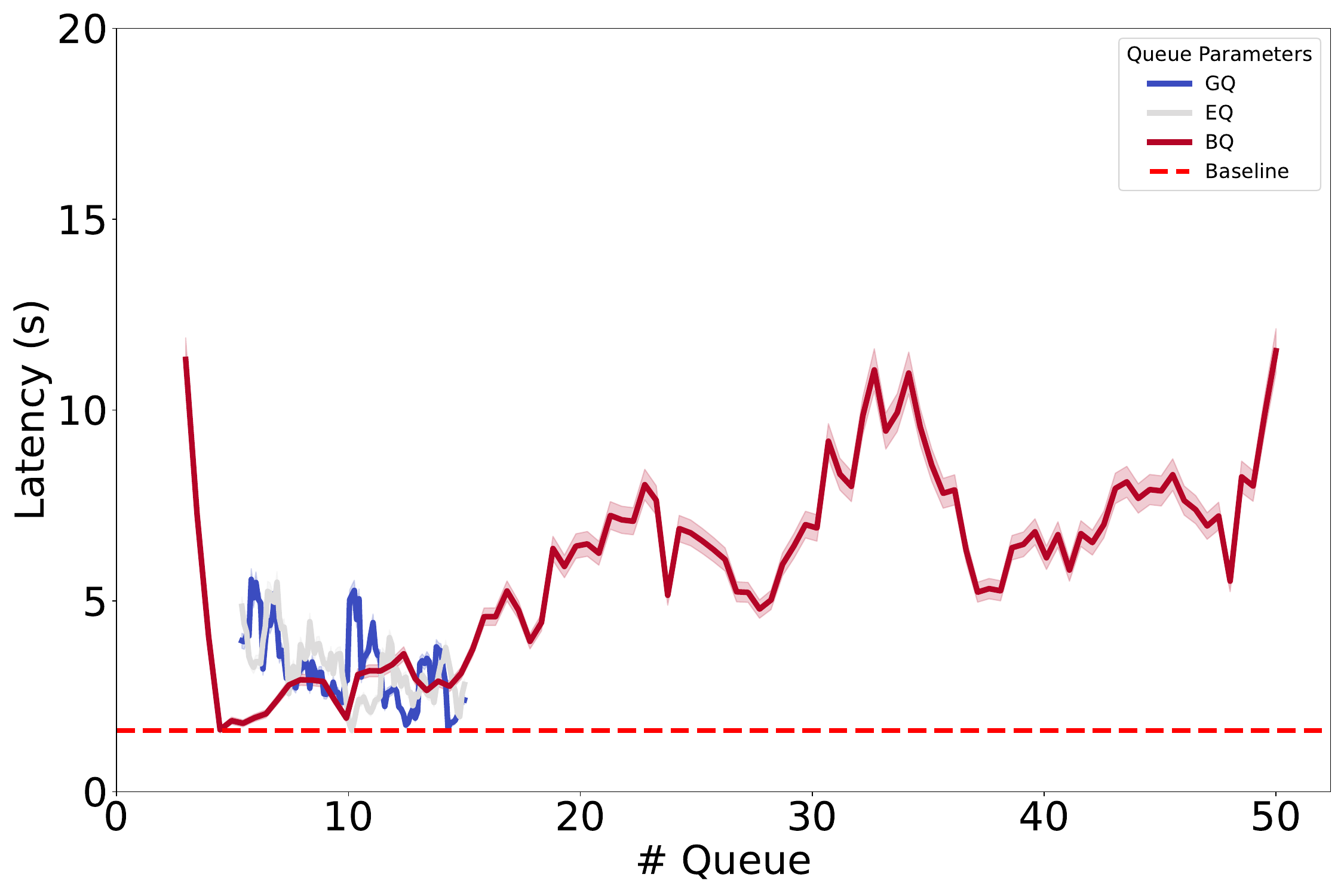}
    \caption{}
    \label{fig:queue_mrt}
  \end{subfigure}
  \hfill
  \begin{subfigure}[b]{0.32\linewidth}
    \centering
    \includegraphics[width=\linewidth]{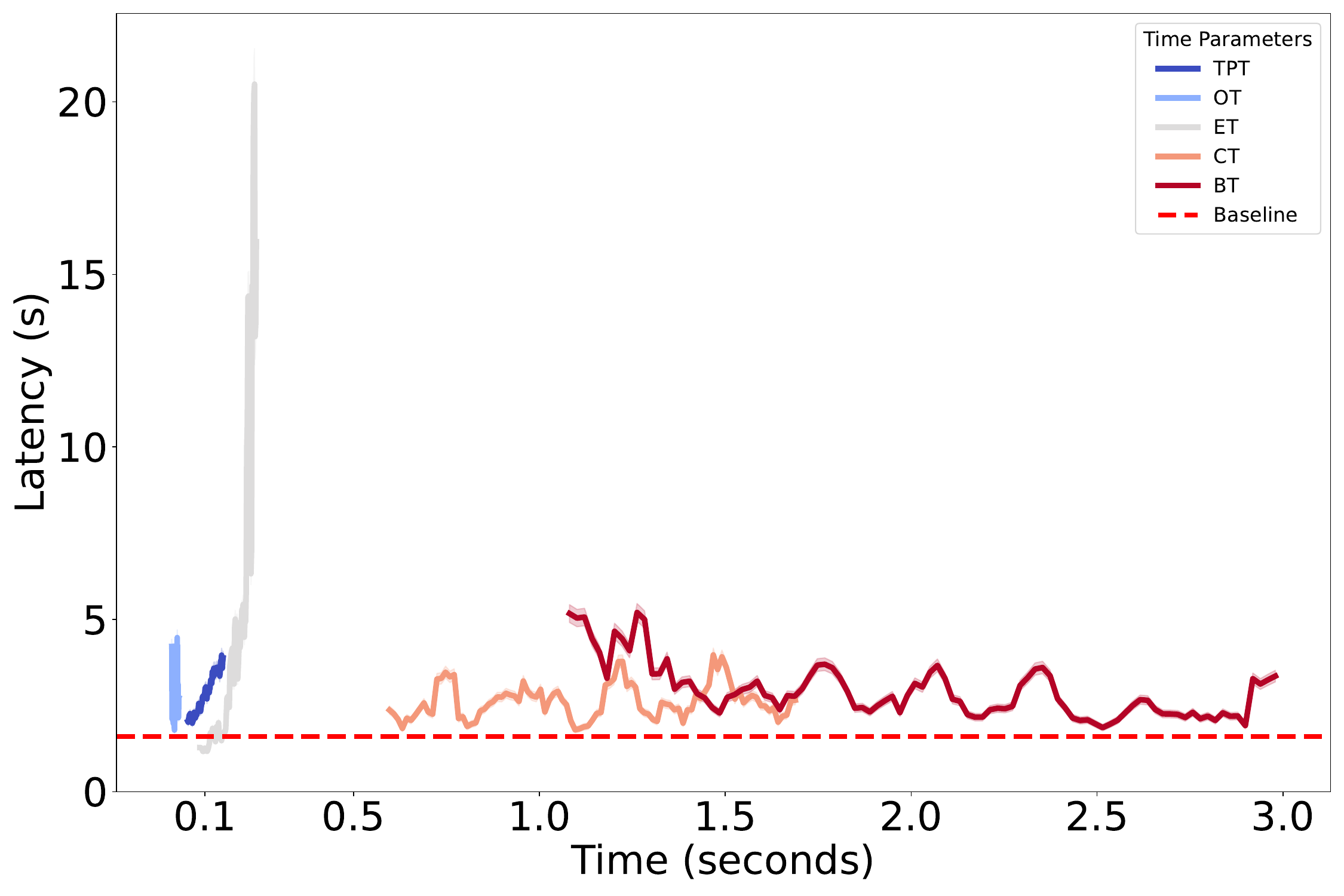}
    \caption{}
    \label{fig:time_mrt}
  \end{subfigure}
  \caption{Impact of Parameters on Latency}
  \label{fig:impact_mrt}
\end{figure*}

Figure \ref{fig:arrival_mrt} shows how the arrival rate influences latency.
It is observed that an increase in the arrival rate results in an increase in latency, from 1110 ms for a rate of 10 tps to 20000 ms for a rate of 100 tps.

Figure \ref{fig:queue_mrt} demonstrates the effect of queue parameters on latency.
Block size (BQ) stands out, as larger blocks may delay filling and processing new blocks.

Figure \ref{fig:time_mrt} highlights the influence of timing parameters on latency.
The endorsement time (ET) has the most significant impact.
Other times, even with variations above or below, have a lesser impact on this metric.

The throughput metric measures the number of transactions carried out over a period of time \cite{thakkar2018performance}.
This paper focuses on the number of transactions persisted on the blockchain per second.
Throughput can be calculated using Equation \ref{eq:t}, with \textbf{E(Place)} representing transactions in \textbf{O1} and \textbf{C1}, and \textit{t(Transition)} indicating the time associated with the \textbf{CT} transition (commit time).

\begin{equation}
\label{eq:t}
    \begin{aligned}
        \textbf{Throughput} = \frac{\text{E(Place)}}{\text{t(Transition)}}
    \end{aligned}
\end{equation}

\begin{figure}[htbp]
\centering
\begin{subfigure}[b]{0.485\linewidth}
\centering
\includegraphics[width=\linewidth]{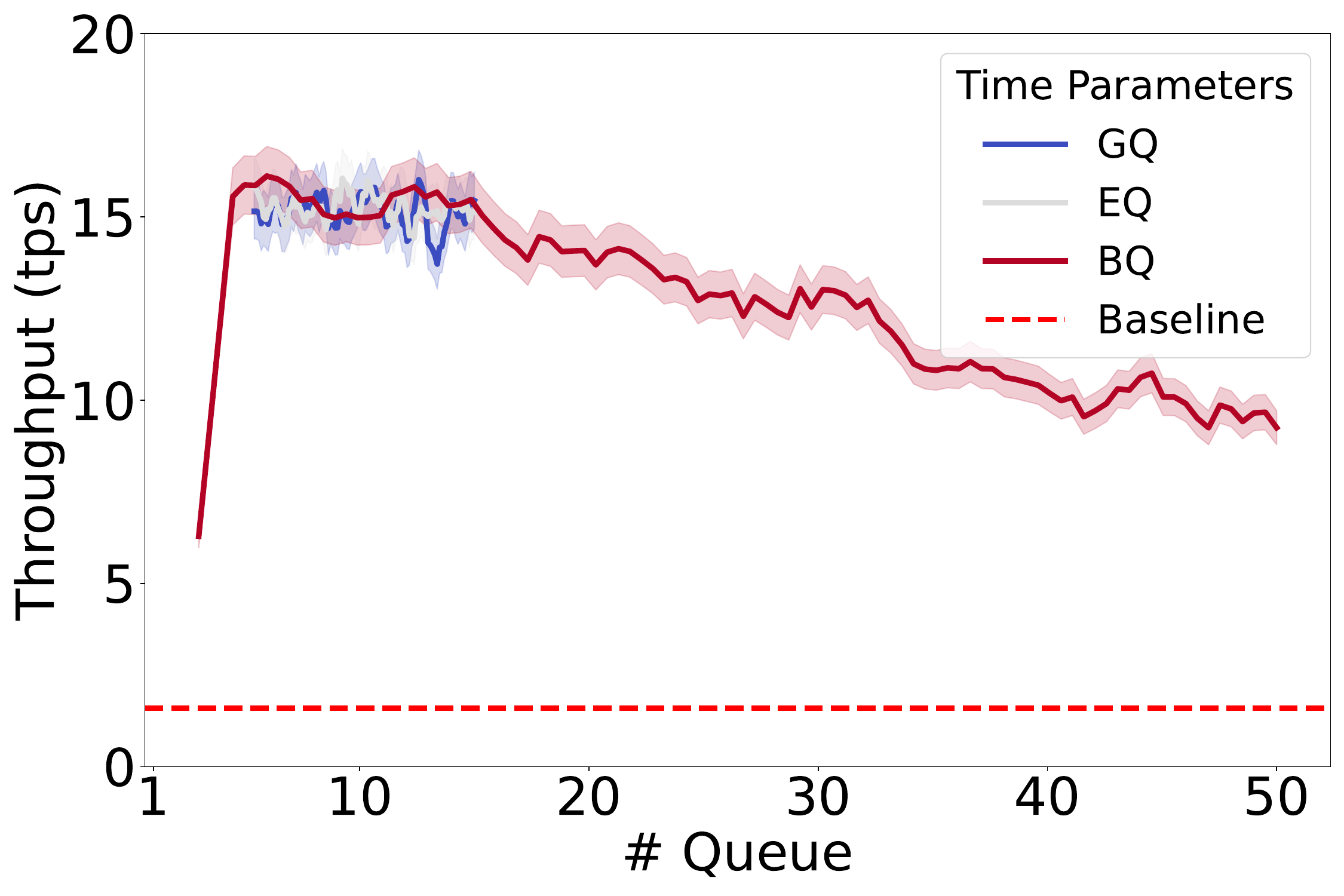}
\caption{}
\label{fig:queue_tps}
\end{subfigure}
\hfill
\begin{subfigure}[b]{0.485\linewidth}
\centering
\includegraphics[width=\linewidth]{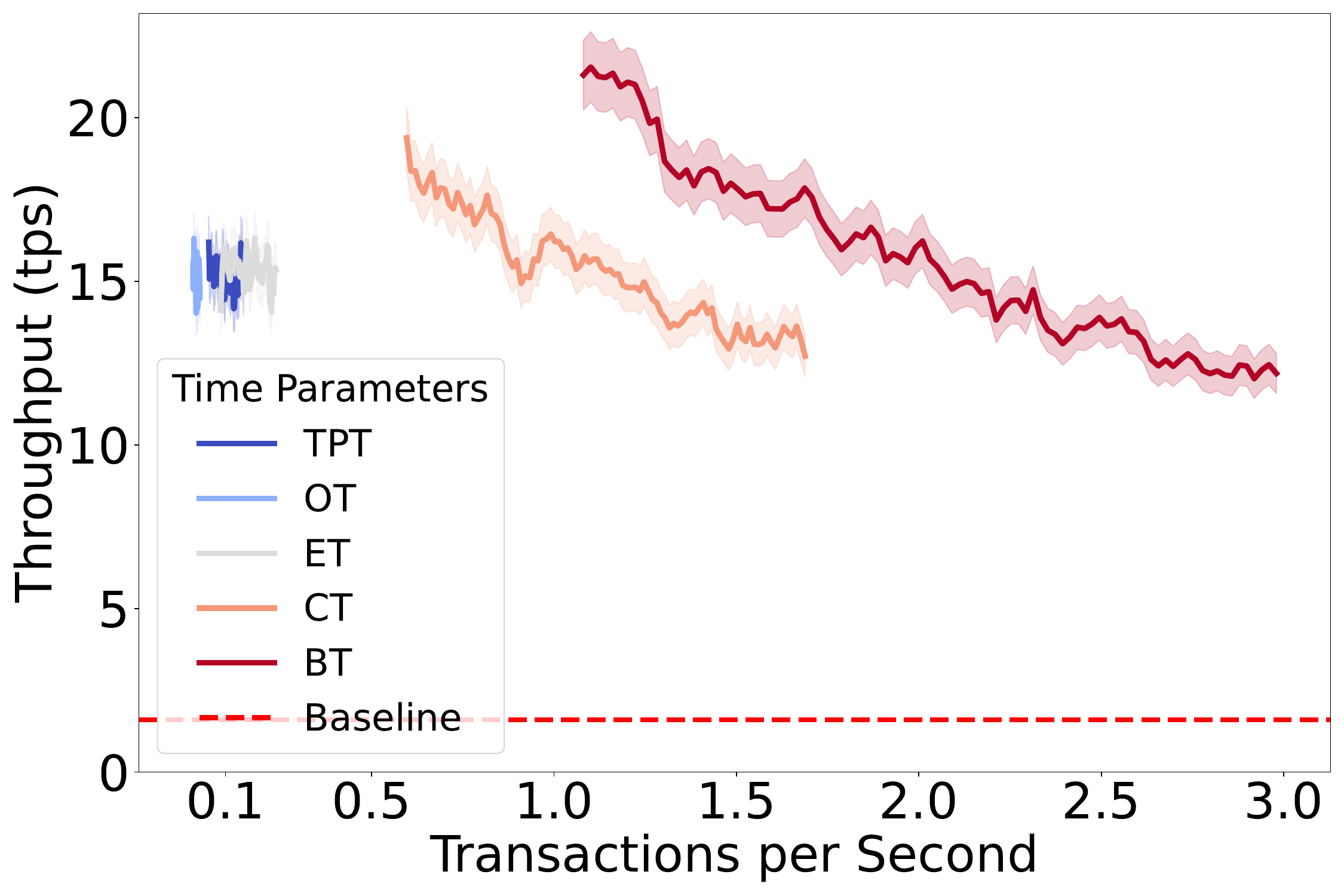}
\caption{}
\label{fig:time_tps}
\end{subfigure}
\caption{Impact of parameters on throughput}
\label{fig:impact_tps}
\end{figure}

The arrival rate constantly impacts the system's throughput.
Meanwhile, Figure \ref{fig:queue_tps} shows that the negative impact of block size (BQ) on throughput is greater than that of other queues in the system, and this is intensified by the maximum time until a block is completed (BT), as can be seen in Figure \ref{fig:time_tps}, the greater the value of BT, the lower the system's throughput.

% \begin{figure*}[htbp]
%   \centering
%   \begin{subfigure}[b]{0.32\linewidth}
%     \centering
%     \includegraphics[width=\linewidth]{img/arrival_tps_iscc.pdf}
%     \caption{Frequency x Throughput}
%     \label{fig:arrival_tps}
%   \end{subfigure}
%   \hfill
%   \begin{subfigure}[b]{0.32\linewidth}
%     \centering
%     \includegraphics[width=\linewidth]{img/fila_tps_iscc.pdf}
%     \caption{Queue x Throughput}
%     \label{fig:queue_tps}
%   \end{subfigure}
%   \hfill
%   \begin{subfigure}[b]{0.32\linewidth}
%     \centering
%     \includegraphics[width=\linewidth]{img/tempo_tps_iscc.pdf}
%     \caption{Time x Throughput}
%     \label{fig:time_tps}
%   \end{subfigure}
%   \caption{Impact of parameters on throughput}
%   \label{fig:impact_tps}
% \end{figure*}

Smaller blocks result in higher throughput, a finding previously confirmed in \cite{silva2023avaliaccao} and now supported by a refined model and a new version of Fabric that includes a gateway and a set of intermediate states that increase precision and improve the accuracy of the obtained results. 
Transactions processed quickly flow more efficiently between endorsement, ordering, and commit.
% A block size of 15 or greater significantly reduces throughput.
% Furthermore, a batch timeout of 2 seconds for very large blocks becomes the only viable path for generating and persisting new blocks.

The utilization metric quantifies the expected number of markings in a specific place \textbf{E(Place)} \cite{maciel2012dependability}, divided by the total capacity of another location directly related to the first, for example, the gateway (G1) and its corresponding queue (GQ). 
The combination of locations in modeling effectively characterizes a resource queue, as shown in Equation \ref{eq:u}.

\begin{equation}
\label{eq:u}
\begin{aligned}
\textbf{Utilization} = \frac{\text{E(Place)}}{\text{Resource Capacity}}
\end{aligned}
\end{equation}

\begin{figure*}[htbp]
\centering
\begin{subfigure}[b]{0.32\linewidth}
\centering
\includegraphics[width=\linewidth]{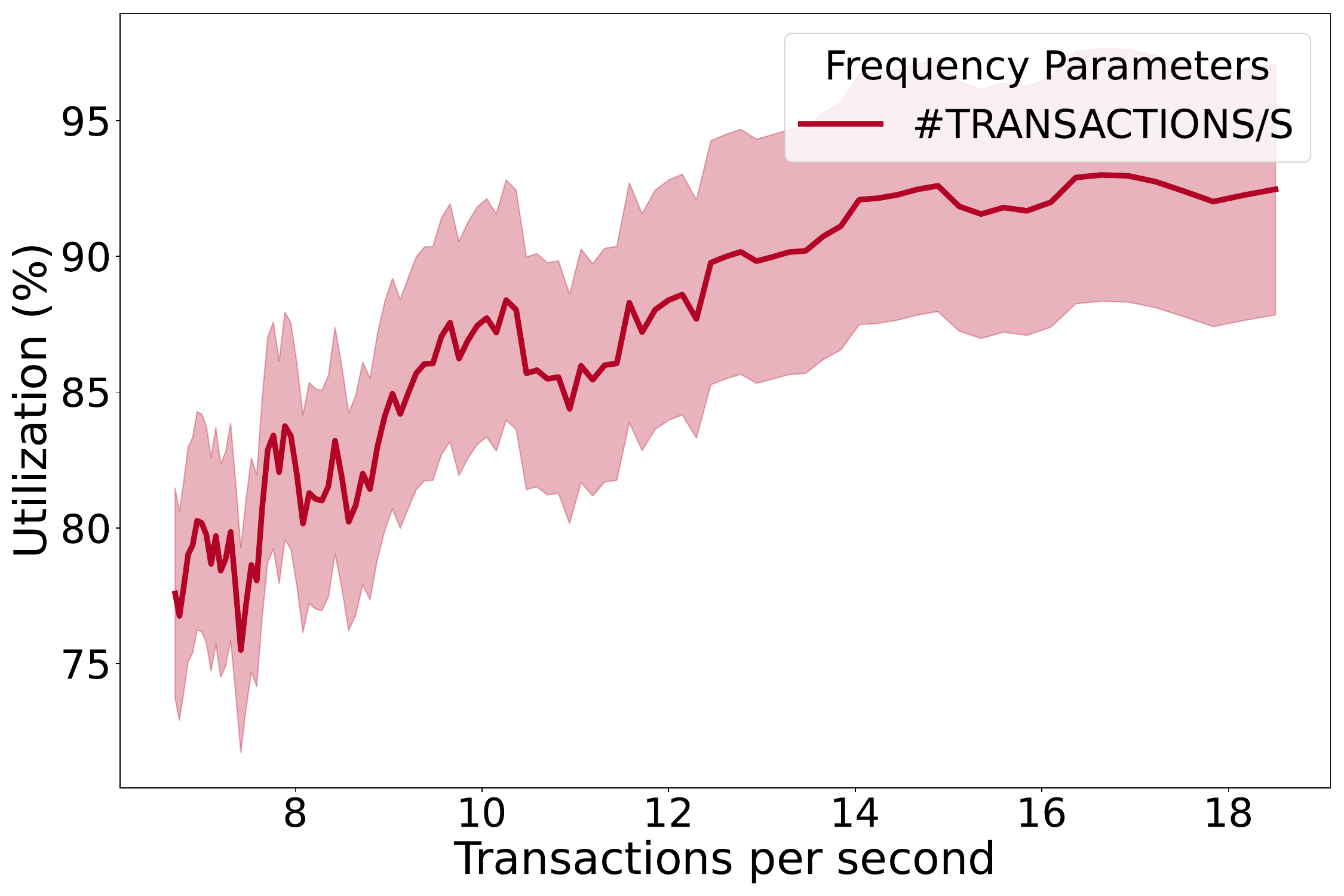}
\caption{}
\label{fig:arrival_utilization}
\end{subfigure}
\hfill
\begin{subfigure}[b]{0.32\linewidth}
\centering
\includegraphics[width=\linewidth]{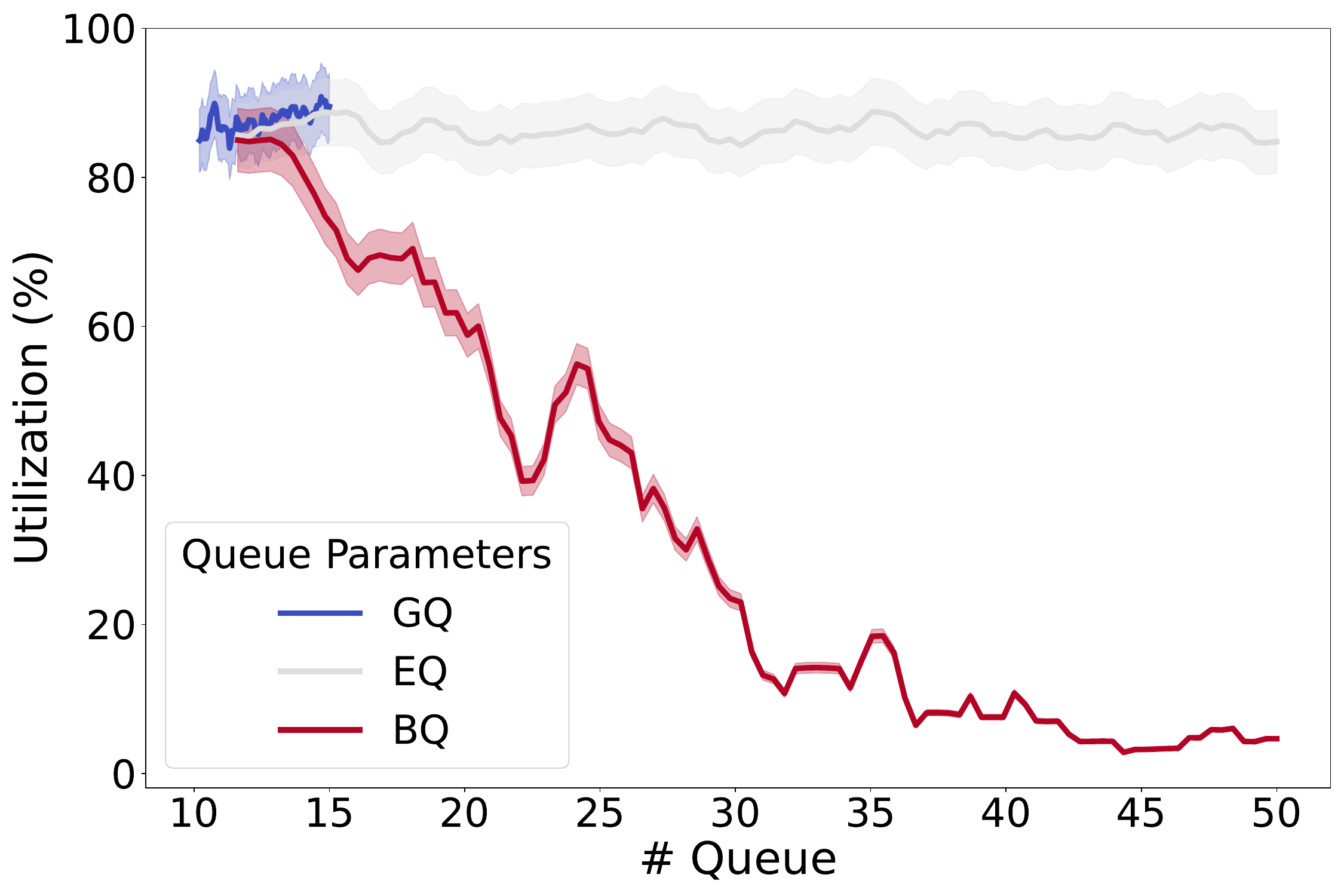}
\caption{}
\label{fig:queue_utilization}
\end{subfigure}
\hfill
\begin{subfigure}[b]{0.32\linewidth}
\centering
\includegraphics[width=\linewidth]{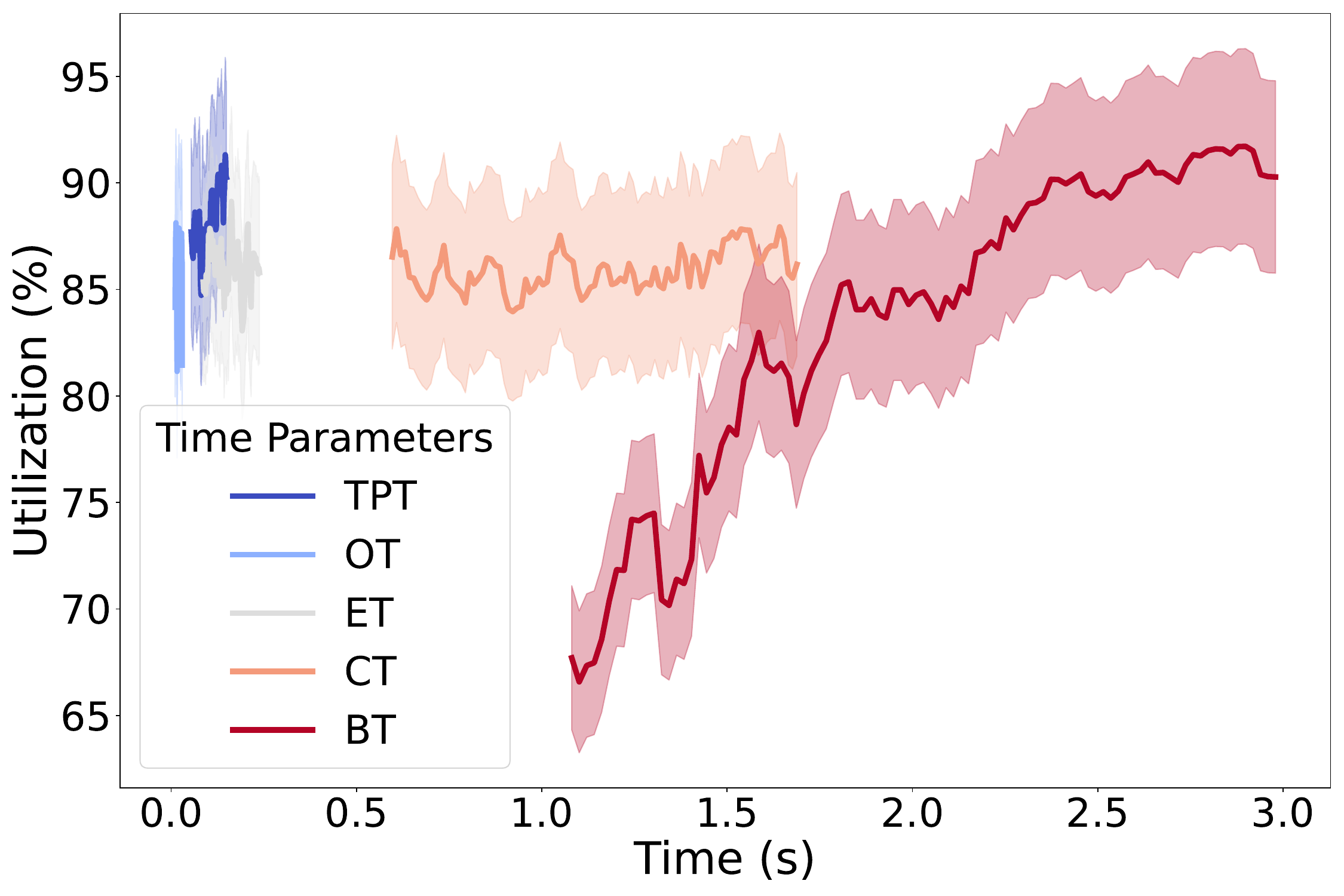}
\caption{}
\label{fig:time_utilization}
\end{subfigure}
\caption{Impact of parameters on gateway utilization}
\label{fig:impact_u}
\end{figure*}

We focus on utilizing the gateway that receives requests from clients and endorsers. As the gateway is a single point of failure, since there is only one gateway in the assessed environment, it is natural to be overloaded. 
Figure \ref{fig:arrival_utilization} shows that the system becomes overloaded with arrival rates below 5 transactions per second, and upon reaching a rate of 20 tps, the gateway utilization reaches 95\%.

Figure \ref{fig:time_utilization} presents the factors related to time and their impact on the utilization. 
The BT has a significant effect; the shorter the time for block formation, the lower the system utilization, which indicates that ordering impacts the gateway more than the arrival of new transactions from the client.

Furthermore, Figure \ref{fig:queue_utilization} confirms the impact of ordering on the gateway by showing that very large blocks (BQ) reduce the system utilization. Therefore, more efficient orderers are a better investment than a larger number of endorsers from the perspective of the evaluated scenario.

\subsection{Endorsement Policy}

The endorsement of a transaction is based on a set of policies described in the source code of that blockchain or in its smart contract. The policies specify how many and which nodes or peers need to agree or attest to the integrity of that transaction through their signature. The three main endorsement policies are \textbf{AND}, \textbf{OR}, and \textbf{K-out-of-N} (KooN).

Assuming we have three endorsing peers. If an AND type endorsement policy is adopted, then all three peers need to endorse the transaction; if it's OR, at least one of the three must endorse it; for KooN, we determine how many peers need to endorse it, for example, \textit{2-out-of-3}, indicates that two out of the three need to do so. Once the transaction is endorsed, it returns to the gateway to be redirected to the orderer.

\begin{figure}[htbp]
\centering
\begin{subfigure}[b]{0.485\linewidth}
\centering
\includegraphics[width=\linewidth]{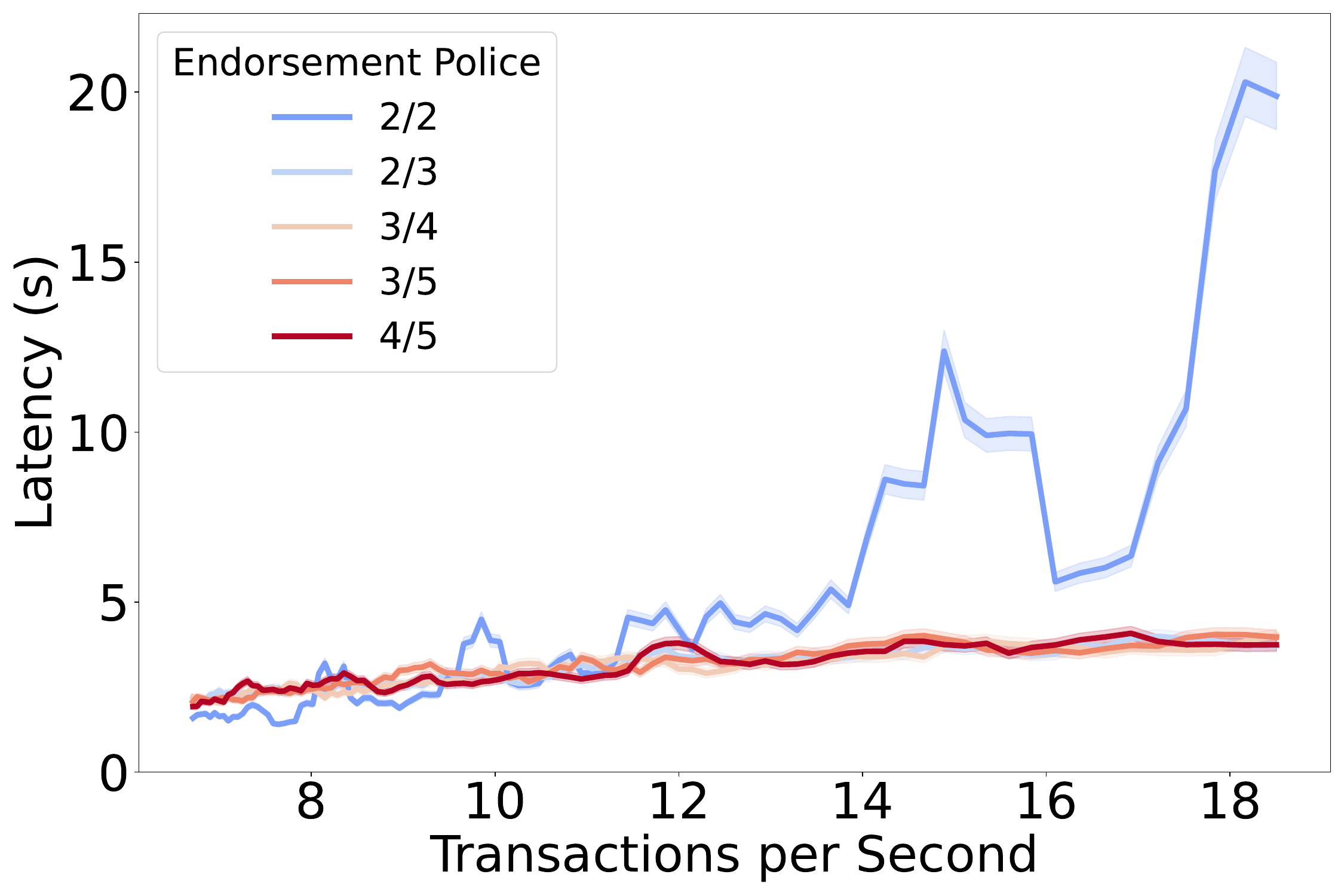}
\caption{}
\label{fig:mrt_police}
\end{subfigure}
\hfill
\begin{subfigure}[b]{0.485\linewidth}
\centering
\includegraphics[width=\linewidth]{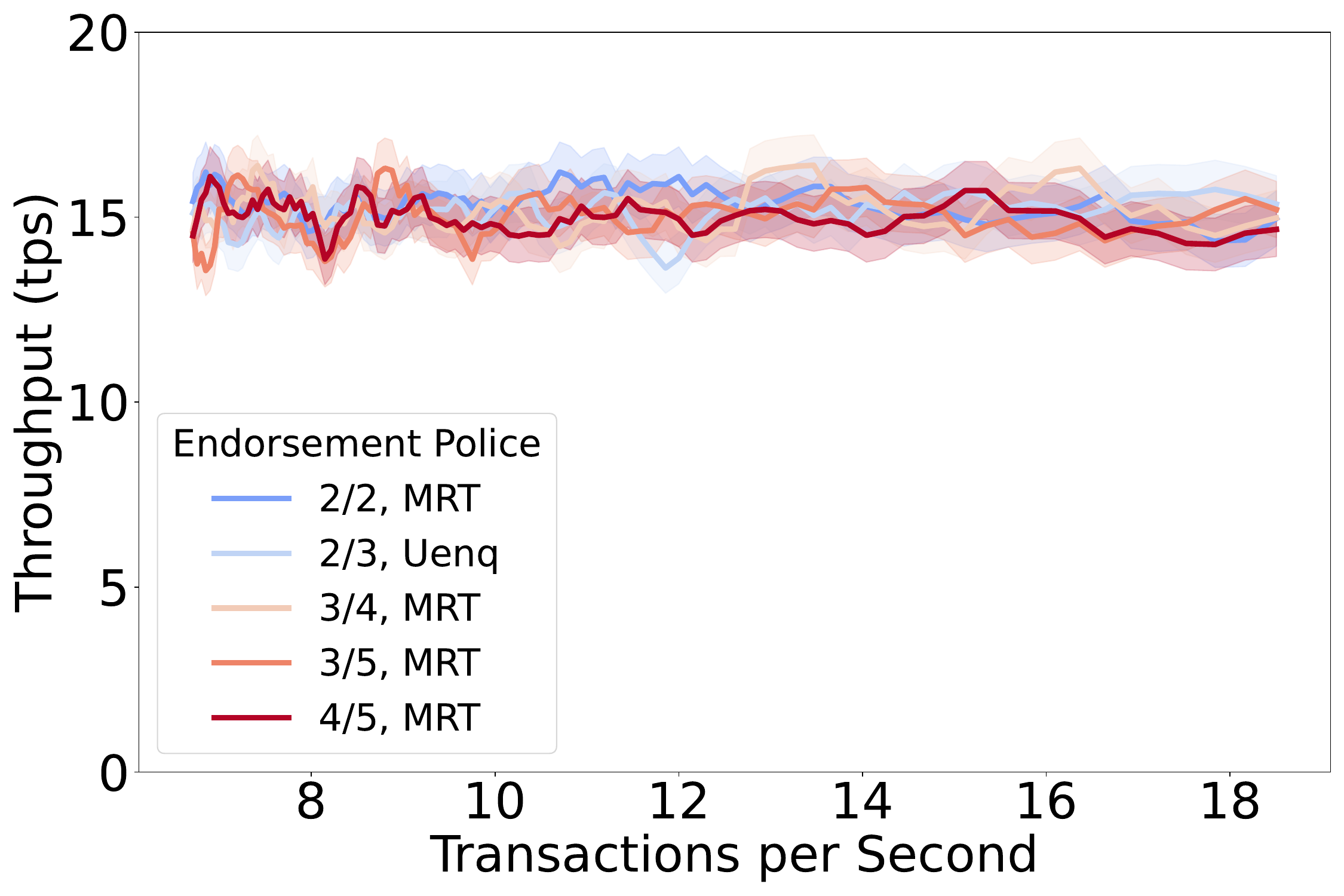}
\caption{}
\label{fig:tps_police}
\end{subfigure}
\caption{Latency and Throughput vs. Endorsement Policy}
\label{fig:polices}
\end{figure}

Figure \ref{fig:mrt_police} shows that adopting a K-out-of-N type endorsement policy provides greater stability to the system in terms of latency and offers extra margin in the availability of the environment since some endorsing nodes may fail, but the service will continue to operate.

Figure \ref{fig:tps_police} shows that adopting endorsement policies has little to no impact on the overall system throughput for the evaluated scenarios. However, it is important to note that the main effect of adopting endorsement policies is on the security of the environment and the relationship between the organizations that comprise it.

\subsection{Distributed Endorsement and Multiple Gateways}

We expanded the scope of the proposed model by introducing a new gateway aimed at reducing overload and mitigating the negative impacts on our single point of failure at the system's entrance. 
Additionally, we considered three geographically distributed endorsers in a \textbf{2-out-of-3} policy by adding latency time to the endorsement process.

This case study aims to assess how the metrics of interest behave with changes in the components that most impacted each of them. For instance, latency, endorsement time, and arrival rate had the greatest impact, while block size was critical for throughput and utilization. We used Table \ref{tab:variation} as the basis for this case study.

The endorsement time for each endorsing peer will correspond to the measured time + network latency based on the distance between the gateway and endorsers. 
The GCPing tool was used and started in Garanhuns, Brazil, and as a destination, the major data center of the Google Cloud Platform (GCP).

The target cities, i.e., the destinations with viable servers chosen based on the shortest distance from the starting point, are São Paulo - Brazil, Santiago - Chile, and South Carolina - USA. The respective latency for each target server was 81 ms, 136 ms, and 197 ms.

\begin{figure}[htpb]
\centering
\begin{subfigure}[b]{0.48\linewidth}
\centering
\includegraphics[width=\linewidth]{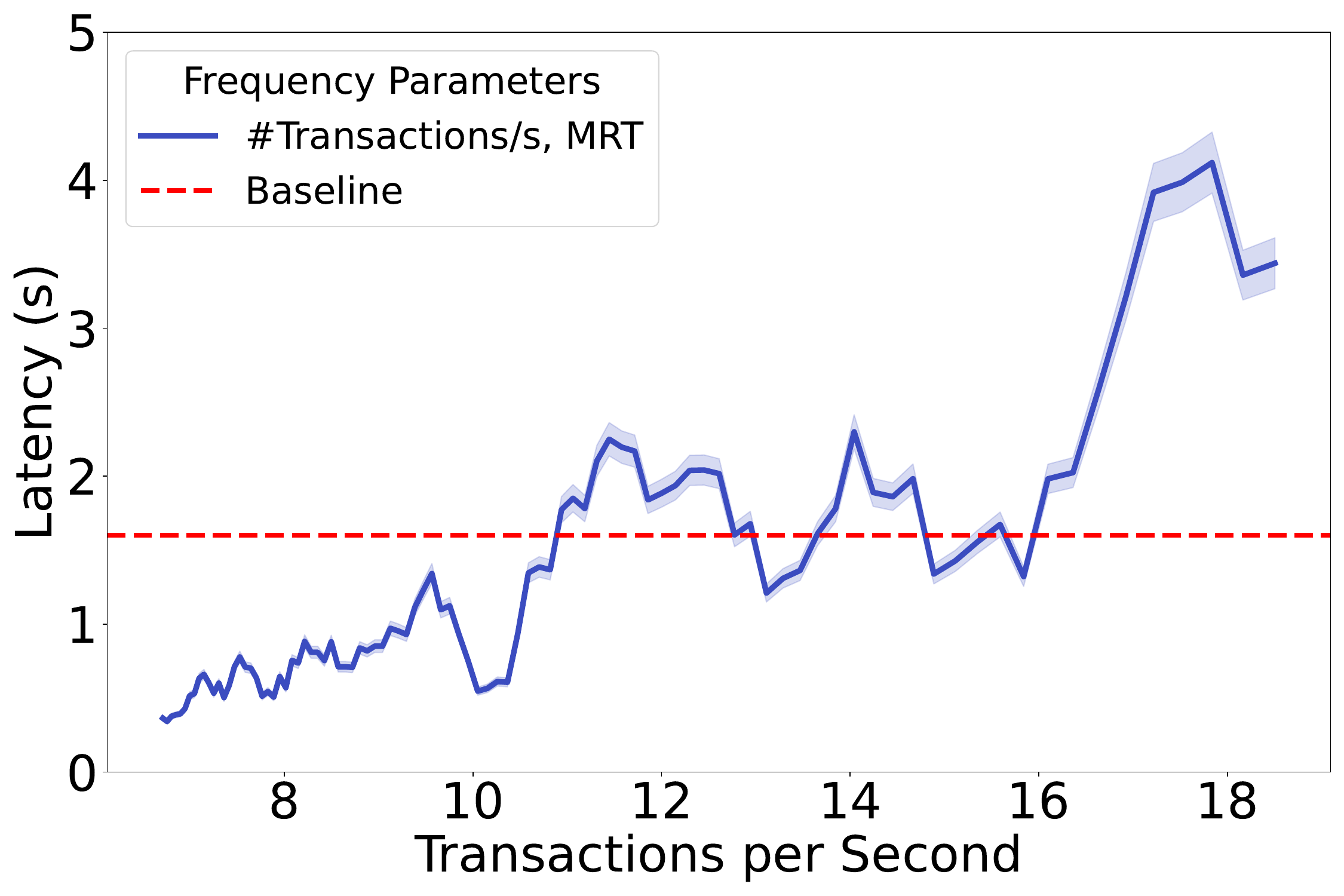}
\caption{}
\label{fig:mrt}
\end{subfigure}
\hfill
\begin{subfigure}[b]{0.48\linewidth}
\centering
\includegraphics[width=\linewidth]{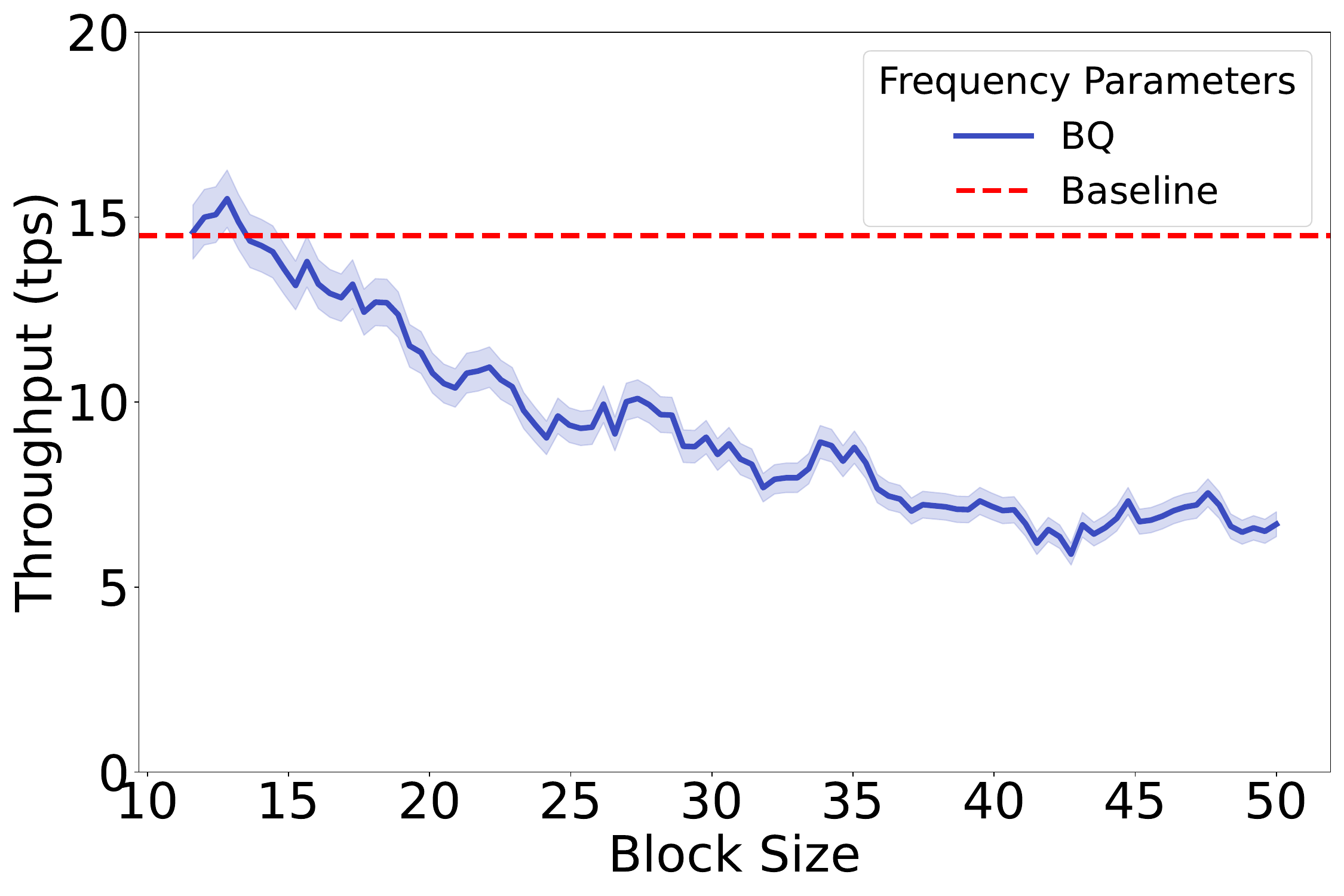}
\caption{}
\label{fig:tps}
\end{subfigure}
\caption{Latency and Throughput vs. Multiple Gateways}
\label{fig:metrics}
\end{figure}

Figure \ref{fig:mrt} presents the impact of the arrival rate on the average response time in an environment with two gateways and three endorsers under the \textbf{2-out-of-3} policy. The previous behavior remains; the higher the arrival rate, the longer the response time. However, the results indicate greater stability. For the first case study, we reached a latency of 20 seconds for an arrival rate of 18 transactions per second.

% For the current case study, we did not exceed 4.5 seconds of latency for the same arrival rate (18 transactions per second), a significant improvement mainly due to the reduction in the overload of the gateways. It is important to note that the gateway scheduling policy considered in the refined model is based on the same probability (50\%) of a transaction going through one gateway or the other.

Figure \ref{fig:tps} presents the impact of block size on throughput in the same environment. 
We noticed the same behavior in the first case study, the larger the block size, the lower the system throughput, as large blocks will use the time to be partially filled and not their completeness. 

%-------------------------------------
\section{Conclusion}
\label{sec:conclusion}
% done
This work presented a detailed analysis of the performance of Hyperledger Fabric, employing measurement and modeling to identify systemic bottlenecks.
The emphasis was on latency, throughput (TPS), and utilization metrics.
The experiments focused on the endorsement, ordering, and confirmation (commit) phases of transactions in the Fabric network.
We also evaluated block size and endorsement queue utilization, concluding that enhancing endorsement capacity can increase throughput and reduce latency.
Finally, we examined the impact of an endorsement policy on the assessed metrics.
An infrastructure with more endorsing nodes and nodes distributed across different locations under a policy where most nodes need to endorse the transaction directly affects the system's latency and throughput.
% We recognize as a limitation the need to refine the temporal distributions used to simulate the stages of the transaction flow, which currently rely on exponential distributions.
Future work will explore other distributions, considering the network's expansion across multiple nodes and latency.
% We aim to enhance the model to address a broader spectrum of metrics and aspects of the platform.

\section{ACKNOWLEDGEMENT}

This work was carried out with financial support from the National Council for Scientific and Technological Development (CNPq) under process PDPG-POSDOC-AUXPE No. 88881.830176/2023-01.

\bibliographystyle{IEEEtran}
\bibliography{main}

\end{document}